# The study of adsorption behaviour of a laser dye incorporated into ultra thin films

S. A. Hussain
Department of Physics, Tripura University; India
Email: sah.phy@gmail.com

Abstract:

This work reports the adsorption behaviour of a highly fluorescent laser dye Rhodamine B (RhB) incorporated into ultra thin films in a long chain fatty acid matrix of Stearic acid (SA). Langmuir Blodgett (LB) method and Layer by Layer (LbL) both the film fabrication techniques are combined to study the adsorption of the dye with increasing number of matrix layers on the substrate and also with increasing time of immersion of substrate with upper surfaces covered by matrix SA. ATR – FTIR investigations supported the interaction and consequent complexation in the complex films. UV-Vis absorption and Fluorescence spectroscopy of the ultra thin film confirms the presence of RhB molecules in the complex films fabricated onto solid substrate. The outcome of this work clearly shows successful incorporation of RhB molecules in ultrathin films without altering the photophysical characteristics of the dye.

1. Introduction:

Ultrathin organic films are recently becoming an era of immense interest mainly focusing on various applications in the field of integrated optics, sensors, friction reducing coatings etc [1-5]. In most of these applications a unique geometrical; arrangement of the constituent molecules with each other and also with the substrate is required to form well defined films with interesting photo physical properties. Langmuir Blodgett (LB) technique although being a common method for mono/ multi layered film preparation, shows interesting results in fabricating films with predetermined alterations of different LB parameters such as deposition surface pressure, pH, sub phase temperature etc. using different amphiphilic molecules in general [6-8]. Recent investigations reveal that certain water-soluble cationic or anionic types of materials when interact electrostatically with the amphiphilic molecules of preformed Langmuir monolayer, adsorption of these molecules in the monolayer results subsequently in formation of complex monolayer [9-15].

Another method for fabricating water soluble molecules in ultra thin films was developed recently based on electrostatic interaction of opposite charges [16-17]. Consecutive alteration of adsorption of the anionic molecule and the cationic molecule on the substrate leads to the formation of multilayer assemblies. This technique is well known as Layer by Layer (LBL) or Self Assembled technique. Therefore it is a challenging job to combine this two techniques in order to fabricate super lattice film with well fascinating characteristics. The formation of insoluble monolayer with ionic amphiphilis at the air water interface and then transfer of mono/multi layered LB films onto a solid substrate is well studied. Our current approach is bi-technical, combining both LB and LbL [18-19]. Using LB technique a stable amphiphilic mono/ multilayer is transferred on solid quartz substrates at a particular surface pressure. Adsorption occurs when this substrate is then immersed into the aqueous solution of a oppositely charged fluorescent dye.

The fluorescent behaviour of organic dyes is highly affected by the relationship that molecules can establish with the neighbouring molecules to prevent forming dimmers or aggregates. The incorporation of fluorescent dye in the restricted geometry usually shows huge deviations from characteristic solution behaviour [20].

To the best of our knowledge water soluble xanthene molecular dye Rhodamine B has not yet been studied in such bi-technical approach in ultra thin organic films. In this paper we investigated the adsorption behaviour of this cationic laser dye Rhodamine B into a long chain fatty acid matrix of Stearic acid (SA).The spectroscopic characteristics of nono/multi layered complex films obtained through this combined technique is studied mainly in the light of UV-Vis absorption Spectroscopy. The dye adsorption and hence complex formation on the solid substrate is confirmed through ATR-FTIR spectroscopic study.

## 2. Experimental

Rhodamine B (RhB) and Stearic acid (SA), purity > 99%, were purchased from Sigma chemical company and were used as received. The purity of the sample was checked by UV – Vis absorption and fluorescence spectroscopy prior to use. The chloroform (SRL, India) used as solvent for SA is of spectroscopic grade and its purity was also checked by fluorescence spectroscopy before use.. Ultra pure Milli – Q (resistivity 18.2 $\Omega$ - cm) water was used for solution preparation and as sub phase. The temperature was maintained at 24 $^0$C throughout the experiment.

For film fabrication with the standard amphiphilic molecule SA in LB method, LB deposition instrument- APEX 2000C is used. To first of all 200 $\mu$l of chloroform solution of SA was spread at the air – water interface of the LB trough. Allowing 15 minutes to evaporate the solvent, the barrier was compressed to obtain the desired surface pressure of 15 mN/m and then vertical Y-type deposition of stable SA monolayer was started by dipping the fluorescence grade quartz substrate at a speed of 5 mm/min with a drying time of 5 min after each lift. Thus mono/ multi layered film of SA is obtained as per requirement. Then the substrate is dipped into the aqueous solution of Rhodamine B of concentration $10^{-4}$M in a beaker for adsorption of oppositely charged molecule in LBL method.

UV-Vis absorption and fluorescence spectra were recorded by UV – Vis absorption spectrophotometer (Lambda 25, Perkin Elmer) and Fluorescence spectrophotometer (LS 55, Perkin Elmer) respectively. For ATR – FTIR measurement the complex film through dye adsorption was fabricated onto a zinc sellenide single crystal substrate. A clean zinc sellenide substrate was used for background measurement. FTIR spectrophotometer (Spectrum 100, Perkin Elmer) was used for ATR – FTIR measurement.

## 3. Results and discussions:
### UV-Vis absorption and Fluorescence Emission Spectroscopy:

Figures 1 and 2 show the UV – Vis absorption and steady state fluorescence spectra of SA – RhB film fabricated on quartz substrate using LB-LbL bi-techniques along with the RhB aqueous solution($10^{-4}$M) and microcrystal spectrum for comparison. For spectroscopic measurement ten layer of LB film have been deposited onto a spectroscopic grade quartz slide and then dipped into aqueous solution of RhB of concentration $10^{-4}$M to study the adsorption behaviour.

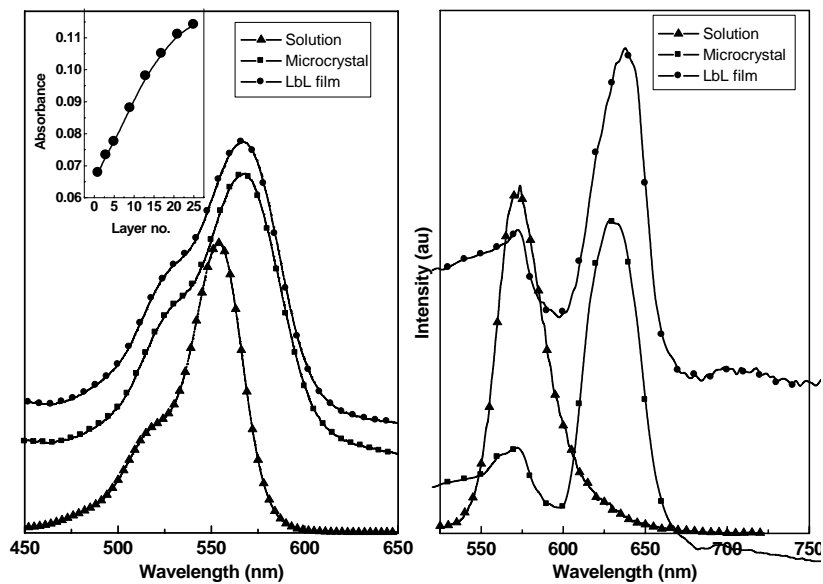

Fig.1.: (a) UV-Vis absorption spectra RhB-SA complex film along with RhB microcrystal and pure RhB solution spectra.

Fig.2.: Fluorescence emission spectra of RhB-SA complex film along with RhB microcrystal and pure RhB solution spectra.

Pure RhB solution spectrum possesses an intense band with peak at 553 nm along with a shoulder at about 515 nm. The 553 nm band is due to the 0 – 0 absorption and trace of monomer, whereas the 515 nm weak shoulder is assigned to be due to the trace of 0 – 1 vibronic transition of monomer [21]. In microcrystal absorption spectrum these two bands are red shifted to 567 and 528 nm respectively. The thin film spectrum also possesses almost similar spectral profile to that of microcrystal.

Observed red shift in the thin film and microcrystal absorption spectra may be due to the change in microenvironment when the molecules are transferred from solution to solid surface. This in-turn affect the electronic energy levels resulting change and band shift in spectrum.

Fluorescence spectrum of RhB aqueous solution shows intense band with peak at 573 nm. An intense longer wavelength band at round 635 nm originates in the RhB microcrystal spectrum and the 573 nm band becomes a weak hump. Fluorescence spectra of SA – RhB thin films show very similar spectral profile to that of microcrystal spectrum. The longer wavelength band originates in the films spectra as the energy levels are affected due to the orientational change in the molecular packing occurred when the molecules are transferred onto solid surface.

The similarity between thin film spectrum to that of the microcrystal spectrum suggest that RhB molecules are incorporated into the complex films and get stacked to form microcrystalline aggregates.

Fig. 3 shows the UV-Vis absorption spectra of RhB adsorbed onto different layered (upto 25 layers) SA LB films. In all cases the immersion time was kept fixed at 30 minutes. It is interesting to note that the absorption spectra of different layered films show almost similar band pattern. The quantitative dependence of adsorption amount of RhB in the SA LB films was determined by monitoring the UV-Vis absorption spectra of RhB adsorbed onto different layered SA LB films. It has been observed that the specific absorbability of the RhB molecules increases with increasing numbers of layers (inset of fig.3). With increasing number of SA layers, it has been observed that the complete adsorption of the RhB molecule requires a longer immersion time. Therefore, the penetration of RhB molecules into SA LB films is a rate determining steps in the adsorption process.

**Time Effect:**

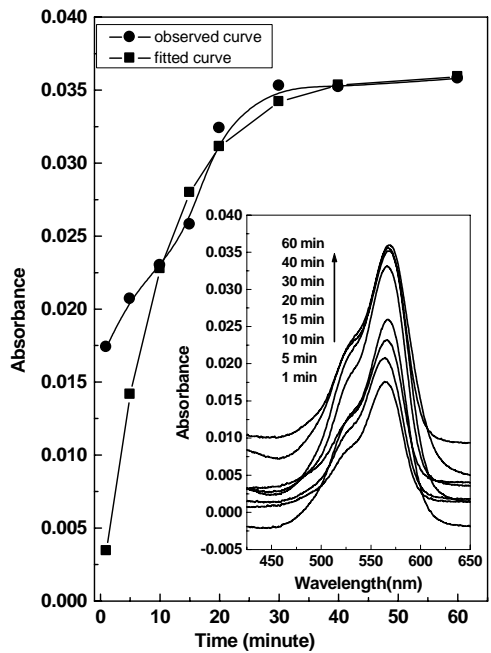

Fig.4: Adsorption kinetics of cationic RhB molecules onto anionic SA LB films as a function of immersion time. Inset shows the UV-Vis absorption spectra of RhB adsorbed SA LB films at varying immersion time.

To monitor the reaction kinetics LB films of SA were deposited onto quartz substrate and this SA film has been immersed into the aqueous solution of RhB ($10^{-4}$M) for different time intervals. In order to quantify the adsorption kinetics, UV-Vis absorption spectra of RhB adsorbed LB films have been measured. Figure 4 shows the adsorption kinetics of cationic RhB molecules onto anionic SA LB films along with the absorption spectra of RhB adsorbed SA LB films at varying immersion time (inset of fig. 4).

From the adsorption kinetics it is seen that the adsorption equilibrium is achieved after almost 30 minutes. It is very interesting to note that initially the absorbance increases very fast however, this rate becomes slow with increase in immersion time. This indicates that the adsorption process becomes slow with increasing time. This is because initially the RhB molecules move towards LB monolayer in a fast diffusive process. After that few RhB molecules on the film surface should change their conformation to accommodate further RhB molecules in films slowing down the adsorption process.

In an attempt to fit the absorbance vs. time data we tried using a single function given in equation (1).

$A = K[1 - \exp(-t/\tau)]$ [1]

Where A is the absorbance taken as proportional to the amount of adsorbed material, K is constant and $\tau$ is the characteristic time. The fitted curve together with the observed curve is shown in figure 4.

It is found that the calculated data are in well agreement to that of the experimental data. This suggests that the adsorption behaviour is of a first order kinetic process with characteristics time 10 minute and k = 0.036.

**Conclusion:**

The results of this work show that water soluble cationic laser dye Rhodamine B can successfully be incorporated onto fatty acid matrix of SA in Langmuir-Blodgett (LB) films through adsorption. Electrostatic interaction occurred between the cationic dye in aqueous solution to the long chain fatty acid molecule in LB films. The presence of RhB molecules in the adsorbed ultra thin films has been confirmed by ATR-FTIR spectroscopy. Reaction kinetics of adsorption has been monitored by UV-Vis absorption spectroscopy. It has been observed that the specific absorbability of RhB molecules increases with increasing numbers of SA layers. However, the penetration of RhB molecules into SA LB films achieves the adsorption equilibrium almost with in a time span of thirty minutes. A comparison between of fitted curve of the reaction kinetics to that of the observed one reveals that the reaction kinetics between RhB and SA films is of first order kinetic process.

**References:**

1. J. D. Swalen, D. L. Allara, J. D. Andrade, E. A. Chandross, S. Garoff, J. Israelachvili, T. McCarthy, R. Murray, R. F. Pease, J. F. Rabolt, K. J. Wynne, H. Yu, Langmuir 1987, 3, 932.
2. A. Ulman, An Introduction to Ultrathin Organic Films: From Langmuir-Blodgett Films of self assemblies, Academic Press, New York, 1991.
3. Special Issue: Organic thin films, Adv. Mater. 1991,3,3.
4. G. G. Roberts, Langmuir Blodgett Film, Plenum Press: New York, 1990.
5. G. Decher, F. Essler, Y. Lvov, J. Phys. Chem. 1993, 97,13773-13777.
6. F. N. Crespilho, V. Zucolotto, O. N. Oliveira Jr., F. C. Nart Int. J. Electrochem. Sci. 1 (2006) 194
7. S. Deb, S. Biswas, S. A. Hussain, D. Bhattacharjee, Chem. Phys. Lett. 405 (2005) 323.
8. S.A. Hussain, P.K. Paul, D. Bhattacharjee, Journal of Colloid and Interface Science 299 (2006) 785.
9. M. Ferreira, C.J.L. Constantino, A. Rlul Jr., K. Wohnrath, R.F. Aroca, J.A. Glacometti, O.N. Oliveira Jr., I.H.C. Mattoso, Polymer 44 (2003) 4205
10. M. Ferreira, R.L. Dinelli, K.Wohnrath, A.A. Batista, O.N. Oliveira Jr., Thin Solid Films 446 (2004) 301
11. S. Biswas, S.A. Hussain, S. Deb, R.K. Nath, D. Bhattacharjee Spectrochimica Acta Part A 65 (2006) 628
12. S. Biswas, D. Bhattacharjee, R.K. Nath, S.A. Hussain Journal of Colloid and Interface Science 311 (2007) 361
13. M. Kawaguchi, M. Yamamoto, T. Kato, Langmuir 14 (1998) 2582
14. J. Engelking, H. Menzel, Eur. Phys. J. E 5 (2001) 87
15. J. Engelking, D. Ulbrich, H. Menzel, Macromolecules 33 (2000) 9026
16. S. A. Hussain, D. Bhattacharjee, Mod. Phys. Letts. B 23 (2009) 3437
17. G. Decher, Science 277 (1997) 1232
18. Y. Lovov, F. Essler, G. Decher, J. Phys. Chem 97 (1993) 13773
19. K. Ray, H. Nakahara, Phys. Chem. Chem. Phys. 3 (2001) 4784
20. B. Leveau, N. Herlet, J. Livage, sanchez, C.Chem. Phys. Lett. 1993, 206, 15.
21. J. Muto, Jpn. J. Appl. Phys. 11 (1972) pp. 1217-1217
22. R. H. A. Ras, C. T. Johnston and R. A. Schoonheydt, Chem. Commun. (2005) 4095 – 4097
23. S. A Hussain, D Dey, S Chakraborty, D Bhattacharjee; Journal of Luminescence 131 (8), 1655-1660
24. PK Paul, SA Hussain, D Bhattacharjee; Journal of Luminescence 128 (1), 41-50
25. D Bhattacharjee, D Dey, S Chakraborty, SA Hussain, S Sinha; Journal of biological physics 39 (3), 387-394